\documentstyle[12pt]{article}
\def\bold#1{\setbox0=\hbox{$#1$}%
      \kern-.025em\copy0\kern-\wd0
      \kern.05em\copy0\kern-\wd0
      \kern-.025em\raise.0433em\box0 }

\def\eea{\end{eqnarray}}
\def\bea{\begin{eqnarray}}
\def\eeas{\end{eqnarray*}}
\def\beas{\begin{eqnarray*}}
\def\ee{\end{equation}}
\def\be{\begin{equation}}
\def\bdm{\begin{displaymath}}
\def\edm{\end{displaymath}}
\def\fr{\frac}

\def\dag{^\dagger}

\def\fpi2{\mbox{F$_\pi$}^2}

\def\mpi2{{m_\pi}^2}
\def\mk{m_K}
\def\mk2{{m_K}^2}

\def\fk2{\mbox{F$_K$}^2}

\begin{document}
\begin{titlepage}
\begin{center}
\hfill TAN-FNT-94-09
\hfill hep-ph/9408360

\vspace*{2.0cm}
{\large\bf ${\bold \Lambda (1405)}$ IN THE BOUND STATE SOLITON MODEL}
\vskip 1.5cm

{\large Carlos L. SCHAT
and Norberto N. SCOCCOLA\footnote[2]{Fellow of the CONICET, Argentina}}
\vskip .2cm
{\large \it
Physics Department, Comisi\'on Nacional de Energ\'{\i}a At\'omica,}\\
{\large \it Av.Libertador 8250, (1429) Buenos Aires, Argentina}
\vskip .2cm
{\large and}
\vskip .2cm
{\large Carlo GOBBI}
\vskip .2cm
{\large \it Center for Theoretical Physics, Sloane Physics Laboratory,}\\
{\large \it   Yale University, New Haven, Connecticut 06511, USA}\\
\vskip .2cm
\vskip 2.cm
August 1994
\vskip 2.cm
{\bf ABSTRACT}\\
\begin{quotation}
The strong and electromagnetic properties of the $\Lambda(1405)$
hyperon are studied
in the framework of the bound state soliton model.
We explicitly evaluate the strong coupling constant $g_{\Lambda^*NK}$,
the $\Lambda^*$ magnetic moment, mean square radii and radiative decay
amplitudes. The results are shown to be in general agreement with available
empirical data. A comparison with results of other models
is also presented.
\end{quotation}
\end{center}
\end{titlepage}

\section{Introduction}

The $\Lambda(1405)$ resonance is one of the most poorly understood
states amongst low-mass baryons. Originally treated as a $KN$ bound
system \cite{Sak60} it was later argued to have a more natural description
in terms of a conventional 3-quark state.
However, in most of the quark model calculations its
rather light mass has been quite hard to describe \cite{IK}.
Within this picture one expects the mass
of the $\Lambda(1405)$
and of the $\Lambda(1520)$ to be very close
since they are ``LS--partners" and LS splittings are known to be
small, at least in the non--strange sector.
Only by introducing further assumptions, like e.g.
three--quark interactions or ``ad hoc" meson--quark interactions~\cite{AMS94}
the $\Lambda(1405)$ mass can be brought into agreement with the empirical
value.
Similar problems are found within bag model calculations \cite{UM89}. In fact,
a cloudy bag model analysis~\cite{VJBT84} seems
to indicate that the $\Lambda(1405)$
is mostly a meson-baryon bound state.

Unfortunately, up to now only
very little empirical information is available about the $\Lambda(1405)$
properties. This situation is very likely to change soon with the anticipated
completion of CEBAF. In fact, one experiment to study the electromagnetic
decays of the $\Lambda(1405)$ using that facility
has been already approved~\cite{CEBAF}.
Proposals to use other planned experimental facilities for the study of
low-lying
excited hyperon properties have been recently put forward as well \cite{LEAR}.

Given this renewed interest in the understanding of the structure of the
$\Lambda(1405)$ (in what follows we will also use the notation $\Lambda^*$) a
comprehensive study of its properties in a soliton model is highly
desirable. The aim of this paper is to calculate the
$\Lambda^*$ strong and electromagnetic properties in the
context of the bound-state soliton model \cite{CK85,SNNR88}.
In this model, the $\Lambda(1405)$ resonance has a natural
explanation as a bound state of a kaon in the background potential of the
soliton. Because of the particular form of the effective interaction
potential the $l=1$ partial wave has a lower bound state energy than the
$l=0$ wave. In fact,  positive parity low-lying hyperons
(e.g. the $\Lambda$ hyperon) are obtained by populating the $P-$wave bound
state
while the $S-$wave soliton-kaon bound system describes
the $\Lambda(1405)$.
The present work also complements previous bound state model studies of strong
coupling constants~\cite{GRS92} and electromagnetic
properties~\cite{KM90,OMRS91,GBR}
where only ground state hyperons were considered.
This will allow a more detailed comparison
with empirical data as well as with other model calculations.

The paper is organized as follows: in Sec. 2 we briefly describe the main
features of the bound state soliton model; in Sec. 3 we discuss the
calculation of the strong coupling constant. Sec. 4 is devoted to the
electromagnetic properties (magnetic moments, mean square radii, radiative
decays). Finally, the conclusions are given in Sec. 5.

\section{General Formalism}

In this section we outline the necessary formalism for the
description of the $\Lambda^*$ properties. As the bound state soliton model
is rather well known by now, we will refer to previous
publications whenever possible.

The effective $SU(3)$ chiral action with an appropriate symmetry breaking term
can be written as
\be\label{lag}
\Gamma =\int d^4 x \Big\{- {f^2_\pi \over 4} Tr(L_\mu L^\mu) +
 {1\over{32 e^2}} Tr [L_\mu, L_\nu]^2 \Big\}
+ \Gamma_{\rm WZ} + \Gamma_{\rm sb}\, .
\ee
Here $\Gamma_{\rm WZ}$ is the non--local Wess-Zumino action
and $\Gamma_{\rm sb}$ is the
symmetry breaking term. Their explicit form can be found,
for instance, in Refs.~\cite{RS91,RRS92}.  In Eq.(\ref{lag})
the left current $L_\mu$ is expressed in terms of the chiral field $U$
as $L_\mu=U^\dagger \partial_\mu U$.

In the spirit of the bound state soliton model, we introduce the
Callan--Klebanov ansatz~\cite{CK85}
\be   \label{ansatz}
U=\sqrt{U_\pi} \, U_{K} \,  \sqrt{U_\pi} \ ,
\ee
where
\bea
U_K \ = \ \exp \left[ i\fr{\sqrt2}{{f_K}} \left( \begin{array}{cc}
                                                        0 & K \\
                                                        K\dag & 0
                                                   \end{array}
                                           \right) \right] \ , \
                                           \ \ \
K \ = \ \left( \begin{array}{c}
                   K^+ \\
                   K^0
                \end{array}
                           \right),
\eea
and $U_\pi$ is the soliton background field written as
a direct extension to $SU(3)$ of the $SU(2)$ field $u_\pi$, i.e.,
\bea
U_\pi \ = \ \left ( \begin{array}{cc}
                       u_\pi & 0 \\
                       0 & 1
                    \end{array}
                               \right ) \ ,
\eea
with $u_\pi$ being the conventional hedgehog solution
$ u_\pi=\exp[i \bold \tau \cdot \hat{\bold r}F (r) ]$.

According to the usual procedure, one expands up to the second order in the
kaon field. The Lagrangian density can therefore be rewritten as a pure
$SU(2)$ Lagrangian depending on the chiral field only and an effective
Lagrangian, describing the interaction between the soliton and the kaon
fields
\be \label{lden}
{\cal L}={\cal L}_{SU(2)}(u_\pi)+{\cal L}(K,u_\pi) \, .
\ee
The explicit form of ${\cal L}(K,u_\pi)$ can be found in
Ref.~\cite{OMRS91}. Therefore in the bound state approach,
the net effect of the harmonic expansion is
the reduction to an effective Lagrangian describing a kaon moving
in the background field of the soliton. The problem now consists in looking
for possible bound states, i.e. in solving the eigenvalue equation
for the meson field $K$ in the potential field of the $SU(2)$ soliton.
The bound state solutions to this wave equation represent stable hyperon
states. Upon a mode decomposition of the kaon field in terms of the
grand spin $\bold \Lambda=\bold L + \bold T$ (where $\bold L$ represents the
angular momentum operator and $\bold T$ is the isospin operator),
the Lagrangian
${\cal L}(K,u_\pi)$ yields a wave equation of the form:
\bea\label{eom}
\left[ -{1\over{r^2}} {d\over{dr}} (r^2 h {d\over{dr}} )
+ m_K^2 + V_{eff}^{\Lambda,l} -
f \ \omega_{\Lambda,l}^2 - 2 \ \lambda \ \omega_{\Lambda,l} \right]
k_{\Lambda,l}(r) = 0.
\eea
Here $\omega_{\Lambda,l}$ is the bound state energy for given $(\Lambda,l)$.
The radial functions $h$, $f$, $\lambda$ and $V_{eff}^{\Lambda,l}$
are functions of  the chiral angle $F(r)$ only. The explicit form of these
functions is given in the Appendix. As customary the chiral angle is
determined by
minimization of the  ${\cal L}_{SU(2)}$ Lagrangian density in Eq.(\ref{lden}).

It has been shown~\cite{SNNR88,RRS92} that for a typical set of
lagrangian parameters there are only two
bound states in the strange sector.
One bound state lies in the $\Lambda=1/2, l=1$ channel
and the other in the $\Lambda=1/2, l=0$ channel.
Since both states have $\Lambda = 1/2$, in what follows
we will label any quantity that depends on the channel quantum numbers
using only the corresponding value of the angular momentum $l$.
{}From the behaviour
of the effective potential at short distances the $l=0$ eigenstate
is expected to lie at higher energy than the $l=1$ state.
Numerical calculations confirm this fact.
Therefore, by populating the $l=1$ state one obtains the
positive parity octet and decuplet hyperons. On the other hand
the $l=0$ state describes the $\Lambda(1405)$ resonance.
Table I shows some typical results for the eigenenergies
together with the definition of Set I and Set II parameters which will be
adopted throughout the paper.

By naively adding once the value of the bound state energy $\omega_l$ to
the soliton mass one obtains only the centroid mass of the $S=-1$ hyperons.
The splittings among hyperons with different spin and/or isospin are given
by the rotational corrections, introduced according to the time--dependent
rotations:
\bea
u_\pi & \to & A u_\pi A^\dagger\, ,    \nonumber \\
K     & \to & A K \, .
\eea
This transformation adds an extra term to the Lagrangian
\be
\delta {\cal L}= {\cal L}_{rot}(u_\pi, K,K^\dagger,A,A^\dagger)\, ,
\ee
which is of order $1/N_c$.
For the particular case of $\Lambda$ and $\Lambda^*$ hyperons
the mass formula which takes into account these rotational corrections
can be written as
\be
M=M_{sol}+\omega_{l}+{3\over {8 \Omega}} c^2_{l}  \, .
\ee
Here, $M_{sol}$ and $\Omega$ are the soliton mass and moment of inertia,
respectively. $c_{l}$ is the hyperfine splitting constant. Its
explicit form for the cases of interest in this paper can be easily
obtained from the general form given  in Ref.~\cite{RRS92}.
The corresponding numerical values are given in Table I.

As reported in Ref.~\cite{RRS92},
positive parity hyperons are well described in the
present model. For example, the calculated mass of the
ground state $\Lambda(1116)$
is 1086 $MeV$ using Set I parameters and 1105 $MeV$ with Set II parameters.
On the other hand the $\Lambda(1405)$ predicted
mass is 1297 $MeV$ using Set I parameters, and 1325 $MeV$ with Set II
parameters.
This results are roughly 100 $MeV$ too low with
respect to the experimental masses.
The situation contrasts with typical quark model predictions,
where the $\Lambda^*$ mass comes out too high in energy.

\section{The Strong Coupling Constant}

Due to its vicinity to the $KN$ system mass, the $\Lambda(1405)$ resonance
plays a dominant role in the analysis of
processes such as $K^- p \to \Lambda \gamma$
and $K^- p \to \Sigma^0 \gamma$. In these analyses the coupling constant
$g_{Kp \Lambda^*}$ is usually considered an adjustable parameter~\cite{WF}.\\
In the bound state soliton model, the $Kp\Lambda^*$ coupling constant can be
explicitly calculated from the effective interaction
Lagrangian upon projection of the
hedgehog solution and of the kaon field onto states of proper spin and isospin.
Since we are interested in the matrix elements between a $\Lambda^*$ state
(rotating soliton--kaon bound system) and a final state composed by a nucleon
(rotating soliton) and a free kaon,
the adiabatic rotation is performed only on the $SU(2)$ hedgehog
and the bound kaon.
Following the general guidelines given in Ref.~\cite{GRS92}
and introducing the identities that relate operators in the
``collective" representation to those expressed in terms of the
conventional spin and isospin representation
\bea
\langle \Lambda^* \vert A^\dagger \vert N K \rangle & = &
{{-i }\over { \sqrt{8 \pi}}} \langle \Lambda^*
\vert -\bold 1 \vert N K \rangle\, ,
\\ \nonumber
\langle \Lambda^* \vert \bold \tau A^\dagger \vert N K \rangle & = &
 {{-i }\over {\sqrt{8 \pi}}} \langle
\Lambda^* \vert -\bold \sigma \vert N K \rangle\, ,
\eea
one obtains the corresponding interaction vertex at threshold:
\begin{equation}
{g_{\Lambda^* K N}\over{\sqrt{4 \pi}}} = {1\over{\sqrt{2}}}
\int dr \ r^2 \ k_0 \left[ f \ m_K\,  \omega_0 +
                           \lambda  (m_K + \omega_0)
                            - m_K^2 - V_{eff}^{0}  \right] \, .
\label{cc}
\end{equation}
As already mentioned, the functions $f$ and $\lambda$  are defined
in Appendix. $V_{eff}^{0}$ can be obtained from the general
expression given in Appendix, replacing the values of
$\Lambda=1/2$ and $l=0$.

At this stage it should be noticed that there is a certain
ambiguity in the definition of the $Kp\Lambda^*$ coupling within our model.
In general, for a negative parity resonance the
pseudoscalar and pseudovector couplings have the form
\bea \label{pssc}
{\cal L}_{\rm PS} & = & g_{\Lambda^*NK} \ \bar u_{\Lambda^*} \ u_N \ K \, ,  \\
\label{psvc}
{\cal L}_{\rm PV} & = &
-i {g_{\Lambda^*NK} \over{M_{\Lambda^*} - M_N} } \
\bar u_{\Lambda^*} \gamma_\mu \ u_N \ \partial^\mu K  \, .
\eea
Integrating by parts Eq.(\ref{psvc}) and using the
free Dirac equation of the $\Lambda^*$ and $N$ one would obtain
Eq.(\ref{pssc}). Therefore both forms of the coupling are
the same for free baryons.
However, if we now perform the non-relativistic
reduction of these interaction
Lagrangians at threshold we get expressions which are somewhat different. From
the pseudoscalar interaction we obtain
\be
{\cal L}_{\rm PS} = g_{\Lambda^*NK} \ \chi^\dagger_{\Lambda^*} \ \chi_N \ K
\, ,
\label{psscnr}
\ee
where $\chi$ are the baryon spinors. On the other hand the non-relativistic
reduction of the pseudovector interaction reads
\begin{equation}
{\cal L}_{\rm PV} =
g_{\Lambda^*NK} \ { m_K \over{M_{\Lambda^*} - M_N} } \
\chi^\dagger_{\Lambda^*} \ \chi_N \ K \, .
\label{psvenr}
\end{equation}
Similar differences are known to happen in the case of the strong couplings
of the ground state $\Lambda$~\cite{Coh87}.
Therefore if only the non-relativistic form of the interaction
Lagrangian is known (as it is the case in the bound state soliton
model) there is no unique definition of the
$g_{\Lambda^*NK}$.
In writing Eq.(\ref{cc}) we have assumed a pseudoscalar coupling,
corresponding to the reduction (\ref{psscnr}). Had we assumed
a pseudovector form for the interaction lagrangian, the r.h.s.
of Eq.(\ref{cc}) would have had to be multiplied by
${M_{\Lambda^*} - M_N \over{m_K}} \simeq 0.94$. Since this factor is quite
close to unity the difference in the numerical results is not
very significant.

Numerical evaluation of Eq.(\ref{cc}) gives
$g_{\Lambda^* NK}=1.6$ for Set I
parameters and $g_{\Lambda^* NK}=2.2$ for Set II.
These values are within the range of typical empirical results
quoted in the literature~\cite{Dum83}. Moreover, in a very recent
analysis of the empirical $KN$ scattering lengths~\cite{LJMR93} the
value $g_{\Lambda^* NK} \simeq 1.9$ has been obtained.
On the other hand, chiral bag model calculations~\cite{UM91,UM93} yield
a smaller value $g_{\Lambda^* NK}=0.46$.

\section{The Electromagnetic Properties}

The electromagnetic properties can be derived entirely from the
electromagnetic current
\be \label{em}
J_\mu^{\rm e.m.} = J_\mu^3+ {1 \over {\sqrt{3}}} J_\mu^8
\ee
obtained from the effective Lagrangian Eq.(\ref{lag})
by means of the Noether theorem.
Once the Callan--Klebanov ansatz is used,
the current is naturally divided into isovector and isoscalar parts.
The first one contributes to the $\Lambda^*$ magnetic moment and mean
square radii. It also describes the $\Lambda^* \to \Lambda \gamma$ decay
($\Delta I=0$). The isovector component
is responsible for the $\Lambda^* \to \Sigma^0
\gamma$ decay ($\vert \Delta I \vert =1$).

\subsection{Magnetic Moments}

The standard expression for the magnetic moment operator reads:
\be
\bold \mu = {1\over 2} \int {\rm d}^3 r \, \bold r \times \bold J^{\rm e.m.}
\, .
\ee
Since the $\Lambda^*$ is an isoscalar resonance,
there is no purely solitonic contribution. Namely, only the term in the
isoscalar piece of the current which is quadratic in the kaon field
contributes.
A straightforward calculation leads to an expression
for the $\Lambda^*$ magnetic moment of the form:
\be
\mu_{\Lambda^*} = {1\over 2}\, \Bigl( c_0
           \, \mu_{s,sol}- a(k_0) \Bigr) \, ,
\ee
where the isoscalar soliton magnetic moment $\mu_{s,sol}$
and $a(k_0)$ are
\bea
\label{msol}
\mu_{s,sol} & = & -{{2 M_N}\over {3 \pi \Omega}}
\int {\rm d}r \, r^2 \sin^2 F F' \, , \\
\label{mkaon}
a(k_0)  & = & {4 \over 3} M_N \int {\rm d}r \, r^2
\Bigl\{ k_0^2 \sin^2 {F\over 2}  \\ \nonumber
               &  & + {1\over { 4 e^2 f_K^2}} \bigl[
 k_0^2 \sin^2 {F\over 2}(F'^2 +{{4 \sin^2 F}\over r^2}) -
   3 k_0 k'_0 \sin F F' \bigr] \Bigr\}\, .
\eea
Here, $M_N$ is the nucleon mass (the magnetic moment is expressed in Bohr
magnetons).
In comparing these contributions
with those corresponding to the ground state $\Lambda$
(see Eqs.(27-32) in Ref.~\cite{OMRS91}) we notice
that the fact that the kaon is bound in the $l=0$ channel modifies
not only the expression of the hyperfine splitting constant but also
the explicit form of $a(k_l)$. This is because although
both states are bound in the same grand-spin channel their
isospin and spatial structure depend on
the angular momentum $l$.

In Table II, the $\Lambda$ and $\Lambda^*$
magnetic moments as calculated in our model are presented.
As usual results are given with respect to the calculated proton magnetic
moment. This is due to the well known
fact that although soliton models predict a somewhat small absolute value
for the proton magnetic moment, they describe magnetic moments ratios
quite accurately~\cite{KM90,ANW83}.  The predictions for the $\Lambda$ magnetic
moment have already been given elsewhere \cite{KM90,OMRS91}
and are quoted here only to serve as reference. We note
that results for the $\Lambda^*$ are somewhat less dependent on the parameter
sets than in the $\Lambda$ case.
We also observe that in contrast
to the case of the ground state $\Lambda$, we predict a small and positive
magnetic moment for the $\Lambda^*$. This is the result of two effects.
On one hand the hyperfine splitting constant is larger in the $l=0$
channel than in the $l=1$ channel. On the other hand, the value of $a(k_0)$
is smaller than that of $a(k_1)$. This second effect
can be understood by noting that although the quadratic contributions
(i.e. first line in Eq.(\ref{mkaon}))
to $a(k_l)$ have roughly the same value in both channels the quartic
contributions are much smaller in the $l=0$ case. Since
in the integrand of the quartic term contributions
the functions that depend on the chiral angle are peaked at short
distances and the $l=0$ wave function is peaked
at a larger radius than the $l=1$ one, this kind of behaviour is
to be expected.

\subsection{Magnetic and Electric Radii}

The magnetic mean square radius of the $\Lambda^*$
can be obtained by integrating the magnetization density of
Eqs.(\ref{msol},\ref{mkaon}) weighted
with an extra $r^2$ and normalized with a proper
coefficient~\cite{KM90,GBR}. The magnetic
radius then reads
\be
\label{rlamstar}
\langle r^2_M \rangle_{\Lambda^*}=
{1\over {2\, \mu_{\Lambda^*}}}
\Bigl(c_0 \langle r^2 \rangle_{s,sol}- \langle r^2
\rangle_{a} \Bigr) \, ,
\ee
where
\bea
\langle r^2 \rangle_{s,sol} & = &  -{{2 M_N}\over {5 \pi \Omega}}
\int {\rm d}r \, r^4 \sin^2 F F' \, ,
\label{rsol}\\
\langle r^2 \rangle_{a} & = &  {4 \over 5} M_N \int {\rm d}r \, r^4
\Bigl\{ k^2_0 \sin^2 {F\over 2} \nonumber \\
              & & + {1\over {4 e^2 f_K^2}} \bigl[
 k_0^2 \sin^2 {F\over 2}(F'^2 +{{4 \sin^2 F}\over r^2}) -
3 k_0 k'_0 \sin F F' \bigr] \Bigr\}  \, .
\label{rkaon}
\eea
The factor ${1\over 5}$ instead of the factor ${1\over 3}$ in
Eqs.(\ref{rsol}--\ref{rkaon}) derives naturally from the normalization of the
magnetic form factor in the limit of zero momentum transfer.
\par
For the isoscalar $S=-1$ hyperons the electric mean square radii
can be simply written as \cite{KM90,GBR}
\be
\langle r^2_E \rangle ={1\over 2} \Bigl[ \langle r^2 \rangle_{\rm B}
- \langle r^2 \rangle _{\cal S} \Bigr] \, ,
\ee
where the elementary radii are defined as
\bea
\langle r^2 \rangle_{\rm B} & = &
   -{2\over \pi}\int {\rm d}r  \, r^2 \sin^2 F \, F' \, , \\
\langle r^2 \rangle_{\cal S} & = & \, 2
\int {\rm d}r \, k^2_l \, r^4 \, [f
\, \omega_l + \lambda ]  \, .
\eea

Numerical values of the calculated magnetic and electric radii are
given in Table II. We observe that the predicted $\Lambda^*$
magnetic radii are much larger that in the $\Lambda$ case.
This can be understood as follows. In the case of the ground state $\Lambda$
there is a partial cancellation between the  $ c_1 \, \langle r^2
\rangle_{s,sol}$ contribution
and the $\langle r^2 \rangle_a$ term. As a result of this the
corresponding magnetic radius is small. In the
$l=0$ channel, however, the extra $r^2$ factor in
the integrand of Eq.(\ref{rkaon}) makes the $\langle r^2\rangle_a$
almost to vanish.
Therefore there is no partial cancellation in the numerator of
Eq.(\ref{rlamstar}).
This fact together with the small value of $\Lambda^*$ magnetic moment
conspires to give a rather large $\Lambda^*$ magnetic radius.

The results for the
electric radii are easier to understand. The only difference
between the $\Lambda$ and the $\Lambda^*$ cases appears in the
strangeness radii $\langle r^2\rangle_{\cal S}$. Being composed by soliton-kaon
system in an excited
state the strangeness radius is expected to be larger for the $\Lambda^*$.
Since this contribution has to be subtracted from the soliton baryon radius
to get the electric radius, it is reasonable
to get a smaller (and even negative) value for the $\Lambda^*$.

\subsection{Radiative Decay Amplitudes}

As already mentioned, the $\Lambda(1405)$ has two electromagnetic decay
modes, namely $\Lambda(1405) \rightarrow \Lambda \gamma$ and
$\Lambda(1405) \rightarrow \Sigma^0 \gamma$. They are related to the
isoscalar and to the isovector part of the e.m. current, respectively.
The decay amplitude corresponding to these processes can be written as
\be \label{decay}
\Gamma=  k  \sum_{m_i, m_f} \sum_{\lambda=\pm 1}
\vert \langle J_f, m_f \vert \hat \varepsilon^*_\lambda (\hat {\bold k})
\cdot \bold J( \bold k) \vert J_i, m_i \rangle \vert^2 \, .
\ee
Here $\bold k$ is the momentum of the emitted photon,
$\hat \varepsilon^*_\lambda$
its polarization tensor and $ \bold J(\bold k) $ the Fourier transform of the
e.m. current $ \bold J(\bold r) $. Also, $k=\vert \bold k \vert$.
As usual, we sum and average over final and initial spin states.

Explicit calculation shows that the relevant matrix elements of the
e.m. current $ \bold J(\bold r) $ can be
written as
\begin{eqnarray}
\bold J^{\Lambda^* H}(\bold r) & = & < \Lambda^* | \bold J | H >
= { i \over{4 \pi}} \left[ g_1^{\Lambda^* H}(r)\,  \bold T +
g_2^{\Lambda^* H}(r)\,  \bold T \cdot \hat {\bold r} \  \hat {\bold r}
\right] \, .
\end{eqnarray}
Here $H$ represents the final hyperon state, namely $H = \Lambda$
for the isoscalar decay and $H= \Sigma^0$ for the isovector one.
The explicit forms of the radial functions $g_1^{\Lambda^* H}$ and
$g_2^{\Lambda^* H}$ are as follows
\begin{eqnarray}
g_1^{\Lambda^* \Lambda} &=&
      \cos F \ \left[ 1 + {1\over{e^2 f_K^2}} (F'^2 +
       {\sin^2 F \over{r^2}} ) \right] \ {k_0 k_1 \over{r}}
\nonumber \\ & &
   + {3\over{4 e^2 f_K^2}} \left[ F' {\sin F \over{r}}
       (k_0' k_1 + k_0 k_1') - F'^2 \cos F {k_0 k_1 \over{r}} \right]\, ,
\\
\nonumber \\
\nonumber \\
g_2^{\Lambda^* \Lambda} &=&
   -\cos F \ \left[ 1 + {1\over{e^2 f_K^2}} (F'^2 +
    {\sin^2 F \over{r^2}} ) \right] \ {k_0 k_1 \over{r}}
\nonumber \\ & &
        -  ( 1 + {1\over{2 e^2 f_K^2}}
            {\sin^2 F \over{r^2}} ) (k_0' k_1 - k_0 k_1')
\nonumber \\ & &
         - {3\over{4 e^2 f_K^2}} \left[ F' {\sin F \over{r}}
         (k_0' k_1 + k_0 k_1') - ( F'^2 \cos F + 2 F' {\sin F \over{r}} )
          {k_0 k_1 \over{r}} \right]
\end{eqnarray}
and
\begin{eqnarray}
g^{\Lambda^* \Sigma^0}_1 & = &
{ 2 \cos F - 1\over 3 } g_1^{\Lambda^* \Lambda}
       + {1\over{3 e^2 f_K^2}} {\sin^2 F\over r}
        \left\{ k_0' k_1'-
     \bigl[ \omega_0 \omega_1 + {7\over4} F'^2 + 2 {\sin^2 F\over{r^2}}
     \bigr] k_0 k_1 \right\}
\nonumber \\ &  &
  - {{N_c} \over {18 f_K^2 \pi^2}}
        \left[ {{\sin 2 F}\over{2 r} }
        (\omega_1 k'_0 k_1 + \omega_0 k_0 k'_1) \right.
\nonumber \\ & & \qquad \qquad \left.
          + F' {{k_0 k_1}\over r }(\omega_0 \cos^2 {F\over 2}
          +\omega_1 \sin^2 {F\over 2} +\omega_0 \sin^2 F) \right] \, , \\
g^{\Lambda^* \Sigma^0}_2 & = &
 {{2\cos F-1 }\over 3} g^{\Lambda^* \Lambda}_2
- {1\over{3 e^2 f_K^2}} {\sin^2 F\over r}
     \left\{ k_0' k_1'-
       \bigl[ \omega_0 \omega_1 +
          {7\over4} F'^2 + 2 {\sin^2 F\over{r^2}} \bigr] k_0 k_1
\right. \nonumber \\ & &
\qquad \qquad \qquad \left.
            + {2\over r} [ k'_0 k_1 \cos^2 {F\over 2}
                        + k_0 k'_1 \sin^2 {F\over 2} ] \right\}
\nonumber \\ & &
    -{N_c \over {18 f_K^2 \pi^2}}
         \left\{ ({{\sin 2 F}\over r }-F')
        (\omega_0 \cos^2 {F\over 2} + \omega_1 \sin^2 {F\over 2})
{{k_0 k_1}\over r} \right.
\nonumber \\ &  & \qquad \qquad \left.
        -{{\sin 2 F}\over{2 r}} (\omega_1 k'_0 k_1 + \omega_0 k_0 k'_1 )
        - \sin^2 F  \bigl( F' \omega_0 -
        {{\sin F}\over r} (\omega_0 - \omega_1) \bigr) {{k_0 k_1}\over r}
\right\} \nonumber \, .  \\
\eea
Given the form of $\bold J^{\Lambda^* H} (\bold r)$ it is easy to get
its Fourier transform $\bold J^{\Lambda^* H} (\bold k)$. We get:
\begin{eqnarray}
\bold J^{\Lambda^* H} (\bold k) & = &
i \left[ f_1^{\Lambda^* H}(k) \, \bold T +
    f_2^{\Lambda^* H}(k) \, \bold T \cdot \hat{\bold k} \
\hat{\bold k} \right] \, ,
\end{eqnarray}
where
\begin{eqnarray}
f_1^{\Lambda^* H}(k) &=&
\int {\rm d} r \ r^2 \left[ g_1^{\Lambda^* H}(r) \ j_0(kr) +
  {g_2^{\Lambda^* H}(r) \over3} \ ( j_0(kr) + j_2(kr) ) \right] \, , \\
f_2^{\Lambda^* H}(k) &=&
- \int {\rm d} r \ r^2  \ g_2^{\Lambda^* H}(r) \ j_2(kr) \, .
\end{eqnarray}
Here, $j_0$ and $j_2$ are spherical  Bessel functions of
zeroth and second order. Combining all these expressions we get
\bea
\Gamma (\Lambda^* \to \Lambda \gamma)&=&
k \vert f_1^{\Lambda^* \Lambda}(k)\vert^2 \, ,
\\
\Gamma (\Lambda^* \to \Sigma^0 \gamma)&=&
k \vert f_1^{\Lambda^* \Sigma^0}(k)\vert^2 \, .
\eea

Following the standard prescription, we take the $k$ of the emitted photon
to be the energy difference between the initial and final hyperon state.
In should be noticed that since in our model the $\Lambda^*$ mass turns out
to be rather small such splittings are somewhat underestimated.
For this reason in Table III we list the results for
both the calculated $k_{calc}$ (obtained as the difference
between the calculated initial and final hyperon masses)
and the empirical one $k_{emp}$. This last one is of course
the difference between the empirical hyperon masses.
Since $f_1^{\Lambda^* H}$ is basically constant in the
relevant range of $k$-values the decay amplitudes turn out
to be roughly proportional to $k$. Also given in Table III
are the results of a quark model (QM) calculation \cite{DHK83},
an MIT bag model (BM) calculation \cite{KMS85} and
cloudy bag model (CBM) calculation \cite{UM91}.
Finally, we have also listed some empirical values obtained
from an analysis of kaonic atoms decays (KA) \cite{BL91}.
We observe that our results are smaller than those of the
quark model and in reasonable agreement with the bag model
predictions. The main discrepancy with the CBM results
appears in the $\Gamma (\Lambda^* \to \Sigma^0 \gamma)$ decay width
which is predicted to be very small in that model.
In general our results are somewhat larger than those
obtained from the empirical analysis of Ref.\cite{BL91}. It should
be noticed however that in that analysis the poorly known strong coupling
constant $g_{\Lambda^* NK}$ appears as input parameter. In Ref.\cite{BL91}
the value  $g_{\Lambda^* NK} = 3.2$ has been taken. A smaller value of this
coupling constant (as obtained in our model) would lead to decay widths
in closer agreement with our predicted values.

\section{Conclusions}

In this work we have studied the strong and electromagnetic properties
of the $\Lambda(1405)$ resonance using the bound state soliton model.
Within this model such hyperon is composed by an $SU(2)$ soliton
and a kaon bound in a $S$-wave. We have found that the
predictions for the strong coupling constant
$g_{\Lambda(1405)NK}$ are within the range of the
empirically (not very well) known values. This contrasts
with the situation in models based on a dominant 3 quark description
of the $\Lambda(1405)$ where much smaller values are obtained \cite{UM93}.
It is interesting to note that in the case of
the ground state hyperons both type of models
predict similar values for the corresponding coupling constants
\cite{GRS92,UM93}.

We have made predictions for the $\Lambda(1405)$ magnetic moment and
electromagnetic
radii. Unfortunately, these magnitudes are quite difficult to determine
empirically.
In any case, our values for the $\mu_{\Lambda^*}/\mu_p$ are in qualitative
agreement with calculations quoted in the literature \cite{WJC90}. Perhaps more
interesting are the predictions for the $\Lambda(1405)$ electromagnetic decay
amplitudes. We have computed both the decay amplitudes corresponding to
the isoscalar process $\Lambda^* \to \Lambda \gamma$ and to the isovector
one  $\Lambda^* \to \Sigma^0 \gamma$. Our results are much smaller
than those of the quark model\cite{DHK83} and in reasonable agreement with
the bag model results of Ref.\cite{KMS85}. On the other hand, they are
somewhat larger than the values extracted from kaonic atom decays \cite{BL91}.
This might be due to the value of $g_{\Lambda(1405)NK}$ used in such analysis.
In any case, it is clear that better empirical information
about the $\Lambda(1405)$ properties is needed.
For this reason, we hope that the
results of the planned experiments for the study of hyperon properties at
CEBAF and other facilities will soon be available.


\vspace*{1.cm}

The authors wish to thank A.O. Gattone, M. Rho and Y. Umino for useful
discussions.
One of us (CG) was partially supported by Fondazione Della Riccia.

\section*{Appendix}

In this Appendix we write down the explicit expressions of the radial
functions appearing in the kaon equation of motion Eq.(\ref{eom}).
These expressions have been derived elsewhere. The effective potential
for a general ($\Lambda$, $l$) partial wave is given by
\bea
V_{eff}^{\Lambda,l} &=& 2 \left( {\sin^2 F/2 \over{r}} \right)^2
\left( 1 + {1\over{4e^2 f_K^2}} \left[ F'^2 + {\sin^2 F\over{r^2}}
\right] \right)
 - {1\over4} \left[ F'^2 + 2 {\sin^2 F\over{r^2}} \right]
\nonumber  \\
       & & - {1\over{4e^2 f_K^2}} \left[ 2 {\sin^2 F\over{r^2}}
(2 F'^2 + {\sin^2 F\over{r^2}} ) \right.
\nonumber \\
              & & \qquad \qquad \left.
- {6 \over{r^2}} \left( {\sin^2 F\over{r^2}} \sin^4 F/2 + {d\over{dr}}
 \left( F' \sin F \sin^2 F/2 \right) \right) \right]
\nonumber \\
& & + \left[ 1 + {1\over{4e^2 f_K^2}}
        \left( F'^2 + {\sin^2 F\over{r^2}}) \right) \right]
 {{l(l+1)} \over r^2 }  \nonumber \\
& & + \left\{ {{4 \sin^2 {F\over 2}}\over r^2 }
\left[ 1 + {1\over{4e^2 f_K^2}}
\left( F'^2 + {\sin^2 F\over{r^2}}\right) \right] \right.
\nonumber \\
& & \left.
- {3 \over{2 e^2 f_K^2}} {1\over r^2} \left[ {{\sin^2 F}\over r^2}
\cos F  - {d\over {dr}}(F' \sin F) \right] \right\}
{{\Lambda (\Lambda + 1) -l (l+1) -3/4}\over 2} \nonumber \\
& & - {f_\pi^2 m_\pi^2 \over{2 f_K^2} } ( 1 -\cos F ) \, .
\end{eqnarray}
The explicit form for the radial functions $h$,
$f$ and $\lambda$ is:
\begin{eqnarray}
h &=& 1 + {1\over{2e^2 f_K^2}} {\sin^2 F\over{r^2}} \, ,
\\
f &=& 1 + {1\over{4e^2 f_K^2}} \left( F'^2 + 2 {\sin^2 F\over{r^2}} \right)
\, ,
\\
\lambda &=& - {N_c\over{8 \pi^2 f_K^2}} {\sin^2 F\over{r^2}} F' \, .
\end{eqnarray}

\pagebreak
{\bf \Large Table and Figure Captions}
\begin{description}

\item [Table 1]: Bound state energies $\omega_l$
and hyperfine splitting constants $c_l$ for
$l=0$ and $l=1$ bound states. The pion mass and decay constant ($m_\pi$
and $f_\pi$) and the Skyrme parameter $e$ are chosen
in order to reproduce the phenomenological values of the $N$ and $\Delta$
baryon masses in the $SU(2)$ sector.
The ratio of the kaon to the pion decay constant is set to its
phenomenological value $f_K/f_\pi \sim 1.22$. For a discussion of the
parameters, see e.g. Ref.~\cite{RRS92} and references therein.

\item [Table II]: Magnetic moments and electric and magnetic mean square
radii (in $fm^2$) of the $\Lambda(1405)$ resonance. We report also the
elementary
contributions together with the
corresponding quantities for the $\Lambda$ hyperon. The results are given
for both Set I and Set II parameters.

\item [Table III]: Radiative decay amplitudes of the $\Lambda^*$
(in $keV$). We quote the results for the empirical and for the calculated
photon momentum. Also listed are predictions of the quark model (QM)
\cite{DHK83},
MIT bag model(BM) \cite{KMS85}, cloudy bag model (CBM) \cite{UM91}
and of an analysis of kaonic atoms decays (KA) \cite{BL91}.

\end{description}

\pagebreak

\begin{center}
{\Large \bf Table I}
\vspace{1.cm}

\begin{tabular}{l|c|c|}
\cline{2-3}
  & SET I & SET II \\
\hline
$\!\!\! \vline \ \, m_\pi$ (input) & 138 $MeV$   &  0           \\
$\!\!\! \vline \ \, f_\pi$ (input) & 54  $MeV$   &  64.5 $MeV$  \\
$\!\!\! \vline \ \,  e $ (input)   &    4.84  & 5.45        \\ \hline
$\!\!\! \vline \ \, \omega_{1}$ ($MeV$)   &    209   &   221       \\
$\!\!\! \vline \ \, \omega_{0}$ ($MeV$)   &    388   &   415       \\ \hline
$\!\!\! \vline \ \, c_{1}$         &   0.39   &  0.50       \\
$\!\!\! \vline \ \, c_{0}$         &   0.78   &  0.77       \\ \hline
\end{tabular}
\end{center}

\begin{center}
{\Large \bf Table II}
\vspace{1.cm}

\begin{tabular}{l|c|c|c|c|}
        \cline{2-5}
      & \multicolumn{2}{|c|}{$\Lambda^*$} & \multicolumn{2}{|c|}{$\Lambda$}  \\
\cline{2-5}
      & Set I & Set II & Set I & Set II \\ \hline
$\!\!\! \vline \ \,  \mu_{s,sol}          \, $ &
   0.73      &   0.56       &   0.73       &   0.56         \\
$\!\!\! \vline \ \,  a(k_l)          \, $      &
   0.27      &   0.26       &   1.35       &   1.06         \\
$\!\!\! \vline \ \,  \mu / \mu_p          \, $     &
   0.08      &   0.09       &   -0.27       &  -0.21      \\ \hline
$\!\!\! \vline \ \, \langle r^2\rangle_{s,sol}     \, $  &
   0.40      &   0.28       & 0.40         & 0.28           \\
$\!\!\! \vline \ \, \langle r^2\rangle_{a}          \, $ &
 -0.02        & 0.01         & 0.37          & 0.22           \\
$\!\!\! \vline \ \, \langle r^2_M \rangle          \, $  &
  1.14       &   1.21       &    0.20      &   0.11        \\ \hline
$\!\!\! \vline \ \, \langle r^2\rangle_{B}          \, $ &
   0.47      &    0.35      &    0.47      &   0.35       \\
$\!\!\! \vline \ \, \langle r^2\rangle_{\cal S}          \, $  &
   0.64      &    0.60      &    0.27      &   0.18         \\
$\!\!\! \vline \ \, \langle r^2_E \rangle          \, $  &
   -0.09      &   -0.12       &   0.10       &    0.09        \\ \hline
\end{tabular}
\end{center}

\begin{center}
{\Large \bf Table III}
\vspace{1.cm}

\begin{tabular}{l|c|c|c|c|c|c|c|c|}
        \cline{2-9}
      & \multicolumn{2}{|c|}{Set I} & \multicolumn{2}{|c|}{Set II}
      & QM & BM & CBM & KA \\
       \cline{2-5}
   & $ k_{emp}$ & $ k_{calc}$ &$ k_{emp}$ & $ k_{calc}$  & & & & \\ \hline
$\!\!\! \vline \ \,  \Gamma({\Lambda^* \to \Lambda \gamma}) \, $ &
   67      &    44      &   56       &   40
    & 143 & 60 & 75 & $27 \pm 8$     \\
$\!\!\! \vline \ \,  \Gamma({\Lambda^* \to \Sigma^0 \gamma}) \, $ &
   29      &    13      &   29       &   17
    & 91 & 18 & 2.4 & $10 \pm 4$ or $23 \pm 7$   \\
\hline
\end{tabular}
\end{center}


\newpage

\begin{thebibliography}{99}

\bibitem{Sak60}
R.H. Dalitz and S.F. Tuan,
Phys. Rev. Lett. {\bf 5} (1959) 425;\\
J.J. Sakurai,
Ann.Phys.(N.Y.){\bf 11} (1960) 1;\\
R.C. Arnold and J.J. Sakurai,
Phys. Rev. {\bf 128} (1962) 2808.

\bibitem{IK}
N. Isgur and G. Karl,
Phys. Rev. {\bf D18} (1978) 4187.

\bibitem{AMS94}
M. Arima, S. Matsui and K. Shimizu,
Phys. Rev. {\bf C49} (1994) 2831.

\bibitem{UM89}
Y. Umino and F. Myhrer,
Phys. Rev. {\bf D39} (1989) 3391.

\bibitem{VJBT84}
E.A. Veit, B.K. Jennings, R.C. Barret and A.W. Thomas,
Phys.Lett.{\bf B137} (1984) 415;\\
E.A. Veit, B.K. Jennings, A.W. Thomas and R.C. Barret,
Phys. Rev. {\bf D31} (1985) 1033.

\bibitem{CEBAF}
D.L. Adams {\it et al.} (CLAS Collaboration),
{\it ``Radiative decays of low-lying hyperons"},
CEBAF Letter of intent (unpublished).

\bibitem{LEAR}
P.G. Harris,
{\it ``A possible experiment to study the radiative decay of excited
hyperons"}, in Proc. of ``Future Directions in Particle and Nuclear Physics at
Multi-GeV Hadron Beam Facilities", ed. by D.F. Geesaman (1993), p.~349.

\bibitem{CK85}
C.G. Callan and I. Klebanov,
Nucl. Phys. {\bf B262} (1985) 365.

\bibitem{SNNR88}
N.N. Scoccola, H. Nadeau, M.A. Nowak and M.Rho,
Phys. Lett. {\bf B201} (1988) 425;\\
C.G. Callan, K. Hornbostel and I. Klebanov,
Phys. Lett. {\bf B202} (1988) 269;\\
U. Blom, K. Dannbom and D.O. Riska,
Nucl. Phys. {\bf A493} (1989) 384.

\bibitem{GRS92}
C.~Gobbi, D.O.~Riska and N.N.~Scoccola,
Nucl. Phys. {\bf A544} (1992) 671.

\bibitem{KM90}
J. Kunz and P.J. Mulders,
Phys. Rev. {\bf D41} (1990) 1578.

\bibitem{OMRS91}
Y. Oh, D.-P. Min, M. Rho and N.N. Scoccola,
Nucl. Phys. {\bf A534} (1991) 493.

\bibitem{GBR}
C. Gobbi, S. Boffi and D.O. Riska, Nucl. Phys. {\bf A547} (1992) 633.

\bibitem{RS91}
D.O. Riska and N.N. Scoccola,
Phys. Lett. {\bf B265} (1991) 188.

\bibitem{RRS92}
M. Rho, D.O. Riska and N.N. Scoccola,
Z. Phys. {\bf A341} (1992) 343.

\bibitem{WF}
R.L. Workman and H.W. Fearing, Phys. Rev. {\bf D37} (1988) 3117.

\bibitem{Coh87}
J. Cohen,
Phys. Rev. {\bf C37} (1988) 187.

\bibitem{Dum83}
O. Dumbrajs {\it et al.},
Nucl. Phys. {\bf B216} (1983) 277.

\bibitem{LJMR93}
C-H. Lee, H. Jung, D-P. Min and M. Rho,
Phys. Lett. {\bf B326} (1994) 14.

\bibitem{UM91}
Y. Umino and F. Myhrer,
Nucl. Phys. {\bf A529} (1991) 713.

\bibitem{UM93}
Y. Umino and F. Myhrer,
Nucl. Phys. {\bf A554} (1993) 593.

\bibitem{ANW83}
G. Adkins, C.R. Nappi and E. Witten,
Nucl. Phys. {\bf B228} (1983) 552.


\bibitem{DHK83}
J.W. Darewych, M. Horbatsch and R. Koniuk,
Phys. Rev. {\bf D28} (1983) 1125.

\bibitem{KMS85}
E. Kaxiras, E.J. Moniz and M. Soyeur,
Phys. Rev. {\bf D32} (1985) 695.

\bibitem{BL91}
H. Burkhardt and J. Lowe,
Phys. Rev. {\bf C44} (1991) 607.

\bibitem{WJC90}
R. Williams, C.-R. Ji and S.R. Cotanch,
Phys. Rev. {\bf D41} (1990) 1449.


\end{thebibliography}
\end{document}